# Exploring the Distributed Video Coding in a Quality Assessment Context

A. Banitalebi[*], H. R. Tohidypour

Digital Multimedia Lab, ECE Dept., University of British Columbia

**Abstract**   In the popular video coding trend, the encoder has the task to exploit both spatial and temporal redundancies present in the video sequence, which is a complex procedure; As a result almost all video encoders have five to ten times more complexity than their decoders. In a video compression process, one of the main tasks at the encoder side is motion estimation which is to extract the temporal correlation between frames. Distributed video coding (DVC) proposed the idea that can lead to low complexity encoders and higher complexity decoders. DVC is a new paradigm in video compression based on the information theoretic ideas of Slepian-Wolf and Wyner-Ziv theorems. Wyner-Ziv coding is naturally robust against transmission errors and can be used for joint source and channel coding. Side Information is one of the key components of the Wyner-Ziv decoder. Better side information generation will result in better functionality of Wyner-Ziv coder. In this paper we proposed a new method that can generate side information with a better quality and thus better compression. We've used HVS (human visual system) based image quality metrics as our quality criterion. The motion estimation we've used in the decoder is modified due to these metrics such that we could obtain finer side information. The motion compensation is optimized for perceptual quality metrics and leads to better side information generation compared to conventional MSE (mean squared error) or SAD (sum of absolute difference) based motion compensation currently used in the literature. Better motion compensation means better compression.

## 1. Introduction

In recent years, the distributed video coding (DVC) paradigm has been under a lot of attention and the subject of extensive research. The main reason behind this fact is the applicability of this new paradigm in the widely used video uplink applications such as wireless video cameras, video conferencing using mobile devices, low-power surveillance applications, visual sensor networks, multi-view video coding and etc. These new applications require low complexity video encoders due to their intrinsic power constraint. This is in contrast to the classical video coding setting in which much of the complex task of coding is performed at the encoder using high complexity motion estimation algorithms while video can be decoded several times using rather simple decoders used for example in home devices.

The first examples of DVC were developed in[3,4]. Although they were developed separately and there are some differences between them, the major trend is the same. Video frames are separated into two groups and each is encoded using a different method. The first group of frames (main frames) is intra coded using a technology such as H.264/AVC and the second group (Wyner-Ziv frames) is encoded via the Wyner-Ziv[2] and Slepian-Wolf[1] coding paradigms. Main frames are recovered at the decoder before the WZ (Wyner-Ziv) frames and play the role of the so called side information.

The Wyner-Ziv coding of video as suggested by [3] and as adopted in this paper, consists of several steps as seen in Figure. 1 the major concern of this paper is the side information generation process, referred to as "HVS-based SI generation".

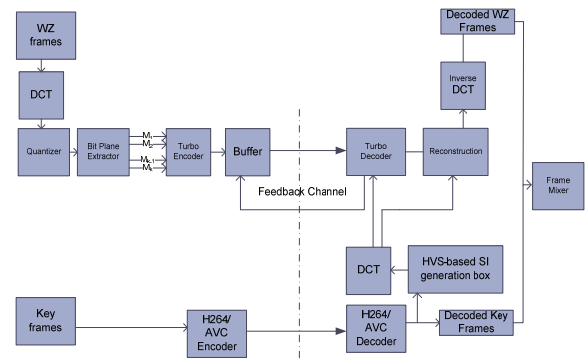

**Figure 1.**   General DVC structure in transform domain.

As a widely known property of the works of [3, 4], the better the quality of side information, the more improvements on rate-distortion performance of the whole system is obtained. The major task of the side information generation process is to produce a frame which resembles the WZ frame in a sense used by the decoder. A lot of work has been done in this regard. The main approach as suggested in [3] is performing a motion-compensated frame interpolation (or extrapolation). This first simple SI generation method consisted of averaging the two nearest key frames and using it as SI (Side Information). To name a few of the methods developed later, Ascendo et al[5] used a block matching method, accompanied by a hierarchical coarse-to-fine bidirectional motion estimation and finally followed by a special motion smoothing process. The authors in [6] used an unsupervised learning approach in a joint work with the decoder. In [7] an adaptive search range was introduced and [8] has also taken some model-based SI generation into account. The authors of [9] suggested a multi-frame SI generation approach using adaptive temporal filtering to estimate the pixel values for SI and motion vector filtering for refinement. The interested reader is referred to [10-17] and the references therein for further reading.

* Corresponding author: dehkordi@ece.ubc.ca (A. Banitalebi)
Email address: htohidyp@ece.ubc.ca (H. R. Tohidypour)

The novelty of this paper is the introduction of a relatively new class of distortion measures in the SI generation process which has never been implemented in DVC architecture before. In this paper we proposed the usage of image quality metrics in SI generation. We explored the use of three most known criterions, SSIM or structural similarity, CW-SSIM or complex wavelet SSIM and VIF or visual information fidelity [18-21]. We used these metrics to modify a technique based on motion compensation of the video at the decoder side to improve the SI for our applications. We showed that by using SSIM, CW-SSIM and VIF measures in generating the side information, better R-D performance can be achieved.

This paper is arranged as follows: In Section 2 we further introduce the concept of distributed source coding and give details on the structure of the DVC setting used in this paper as shown in Fig. 1. Section 3 is dedicated to the definition and application of distortion measures based on human visual system, mostly SSIM, CW-SSIM and VIF. In section 4 we give details about the utilized dataset and section 5 describes the algorithm used in this paper for improvement of SI generation process and further clarifies our contribution. Section 6 presents the simulation results and related talks while section 7 concludes the paper and proposes the future work.

## 2. Distributed Video Coding Basics

Distributed source coding originates in the landmark work of Slepian and Wolf [1]. This theory states that the rate needed for separate encoding of two correlated random variables $X, Y$ is same as the rate for their joint encoding, with arbitrarily vanishing small probability of decoding error as block size goes to infinity and provided that joint distribution, $P_{xy}(x,y)$ is already known at the encoder and stated formally as in (1)

$$R_X \geq H(X|Y)$$
$$R_Y \geq H(Y|X)$$
$$R_X + R_Y \geq H(X,Y) \qquad (1)$$

This notion was later generalized to the case where a distortion metric is also taken into consideration. It was shown in [2] that the achievable rate-distortion pair is given by $R(D) = Inf\{I(X;U|Y)\}$ where $Y \rightarrow X \rightarrow U$ is a Markov chain and there exists a function such as $\hat{X}(u,y)$ where $E\{d(X,\hat{X})\} \leq D$. These works were ignored for a period of almost thirty years and it was not until recently that practical implementations for WZ setting were developed. It is a well-known fact that efficient scalar quantizers and Slepian-Wolf coders can be designed to achieve performances as close to WZ bound as 1.53 dB. It was a brilliant idea to use the WZ coding structure in video coding and the introduction of DVC as done by [3],[4]. Therefore the DVC coding paradigm should use the "quantizer followed by entropy coding" structure of classic video coding but this time specialized for WZ setting. Although the quantizer can be specialized for WZ setting, usually a uniform quantizer is used and the bulk of conditional entropy coding is therefore switched to the Slepian-Wolf code. If work is preferred to be done in the transform domain, DCT transform is usually used before the quantization of data. The quantization indices are Turbo encoded and the parity bits are punctured and sent to the decoder. At the decoder, using various techniques discussed in section V, some side information is generated to be interpreted as a noisy version of the WZ frame transmitted to the decoder. The quantization indices are the result of decoding the side information for the given parity received from the encoder and finally these indices are reconstructed using a predefined reconstruction function. The details as used in the current work are given below.

First, the image is divided into $N \times N$ blocks where usually $N = 8$. A 2D DCT transform is applied to each block and the DCT bands are separated. For each band, the coefficients of the DCT transform are fed to a uniform quantizer. The quantization indices are then extracted to their bit-planes and each bit-plane is given to the Turbo encoder successively from the highest to the lowest bit plane. The turbo encoder which plays the role of the Slepian-Wolf encoder generates the essential parity bits and sends them to the decoder. At the receiver, the Turbo decoder treats the side information as a noisy version of the original WZ bits and using the received parity bits recovers the encoded bit stream. In case of a decoding failure after pre-defined number of iterations, the decoder asks the encoder, through the feedback channel for more parity bits. The initial rate can be set to a minimum rate predefined by several methods such as offline training process or performing some simple side information estimation at the encoder which is not the object of this paper. This procedure is repeated until the bit stream is decoded successfully or a maximum number of retransmissions are reached. The decoded bit stream is then fed to the reconstruction function and reconstruction points, which correspond to pixel values, are then declared. After a WZ frame $I_n$ is fully decoded, it is mixed with previous and next frames, namely $I_{n-1}, I_{n+1}$ to preserve the order of frames.

## 3. Perceptual Quality Metrics

Due to its simplicity, MSE (mean square error) has been the dominant quantitative performance metric in the field of signal processing for many years. But it is shown that there is a big lack of accuracy in MSE when dealing with perceptually important signals such as speech, images and video signals. For those applications newly developed perceptual quality metrics are used.

The MSE and PSNR between two 8-bit image signals x and y are defined as:

$$MSE = \frac{1}{N}\sum_{i=1}^{N}(x_i - y_i)^2 \quad (2)$$

$$PSNR = 10\log\frac{255^2}{MSE} \quad (3)$$

First we explain the SSIM as a popular and state of the art quality assessment metric. We mainly use the primitive formulation of the Wang and Bovic [19,20] in which, the target application is quality assessment of images. SSIM is a function of luminance, contrast and a local structure function of images x and y. Local structural similarity is usually formulated as[19]:

$$S(x,y) = l(x,y).c(x,y).s(x,y)$$
$$= (\frac{2\mu_x\mu_y + C_1}{\mu_x^2 + \mu_y^2 + C_1}).(\frac{2\sigma_x\sigma_y + C_2}{\sigma_x^2 + \sigma_y^2 + C_2}).(\frac{\sigma_{xy} + C_3}{\sigma_x\sigma_y + C_3}) \quad (4)$$

where $\mu_x$ and $\mu_y$ are the local sample means of $x$ and $y$. $\sigma_x$ and $\sigma_y$ are the local sample standard deviations of x and y and $\sigma_{xy}$ is the sample cross correlation of x and y after removing their means. The items $C_1$, $C_2$, and $C_3$ are small positive constants that stabilize each term, so that near-zero sample means, variances, or correlations do not lead to numerical instability. The entire SSIM metric between the original and the reference image is calculated by averaging the local SSIM all over the image.

The application of SSIM in process of searching for the best matching block in motion estimation of the video frames shows a better performance over MSE. We utilized this idea and generated the SI data from a motion compensation technique that uses SSIM as comparison criterion for block matching. Although the application of SSIM shows better performance (better SI and thus better compression), but we go further to improve its performance. We explored the usage of CW-SSIM and VIF. A brief overview of both is given below.

As we see above, SSIM measures the quality by comparing the structures in images. As we know, small geometric distortions are not structural, so a new wavelet domain version of SSIM was introduced to overcome these artifacts [21]. CW-SSIM is usually formulated as [21]:

$$\tilde{S}(c_x,c_y) = \tilde{m}(c_x,c_y).\tilde{p}(c_x,c_y) =$$
$$\frac{2\sum_{i=1}^{N}|c_{x,i}||c_{y,i}| + K}{\sum_{i=1}^{N}|c_{x,i}|^2 + \sum_{i=1}^{N}|c_{y,i}|^2 + K} \cdot \frac{2|\sum_{i=1}^{N}c_{x,i}.c^*_{y,i}| + K}{2\sum_{i=1}^{N}|c_{x,i}.c^*_{y,i}| + K} \quad (5)$$

Where in the previous equation, in the complex wavelet domain, $c_x = \{c_{x,i} | i=1,2,...,N\}$ and $c_y = \{c_{y,i} | i=1,2,...,N\}$ are respectively two sets of wavelet coefficients from the same spatial location in the same wavelet subbands of the two images being compared. K is a small positive constant used for stabilizing. $\tilde{m}(c_x, c_y)$ Is the SSIM index applied to the magnitude of the coefficients and $\tilde{p}(c_x, c_y)$ is calculated by monitoring the difference between phases of $c_x$ & $c_y$. First, CW-SSIM is calculated for each subband of the wavelet decomposition and then average of these values yields an overall CW-SSIM metric for the entire image. More details are available in [21].

VIF quantifies the similarity of two images using a communication framework. It attempts to relate the signal fidelity to the amount of the information that is shared between the two signals, namely the original and the noisy or distorted version. This shared information is quantified using the concept of mutual information which is widely used in information theory. Suppose that we are to compare the quality of two signals such as two images, where one is a reference signal and the other is a noisy version. In the current motion estimation application, one signal is a block of pixels from reference frame and second is a block of pixels from next frame. Let us denote the reference signal by C and the distorted one by D. E is the perceived version of the source signal C by the neurons of the HVS or the Human Visual System and F is perceived version of D. We can write the following equations [18]:

$$d = gc + v, e = c + n, f = d + n \quad (6)$$

In these equations, c and d are random vectors extracted from the same location of the same wavelet subband in the reference and distorted images respectively; g is a scalar deterministic gain factor and v is an independent additive zero-mean white Gaussian noise. This model is a general model but has shown good performances almost everywhere; see [18, 19] for more details on this topic. At the receiver, n is used to model the visual distortion as a stationary zero-mean additive white Gaussian noise process in the wavelet transform domain. The reference image is modeled by a wavelet domain Gaussian Scale Mixture (GSM), which is a good model for natural images [22]. Then c can be modeled as $c = \sqrt{z}u$ where u is a zero-mean Gaussian vector and $\sqrt{z}$ is an independent scalar random variable. According to [18] the VIF is computed by[18]:

$$VIF = \frac{I(C;F|z)}{I(C;E|z)} = \frac{\sum_{i=1}^{N}I(c_i;f_i|z_i)}{\sum_{i=1}^{N}I(c_i;e_i|z_i)} \quad (7)$$

Where *i* is the index of local coefficients patches, with all subbands include. More information about VIF can be found in [18, 19].

## 4. Data

The video frames we used in the simulations are taken from famous "Foreman" and video sequence that was obtained from [27]. The formats of the frames are standard QCIF with 15 HZ frame rate. Figure. 2 show some sample frames used in the simulation. The top frames are frames $I_{n-1}, I_{n+1}$ respectively which are intra coded by the H264/AVC encoder. The bottom left frame is the WZ frame which is compressed by the Turbo encoder and finally the bottom right frame is the side information which is generated by the proposed method of this paper.

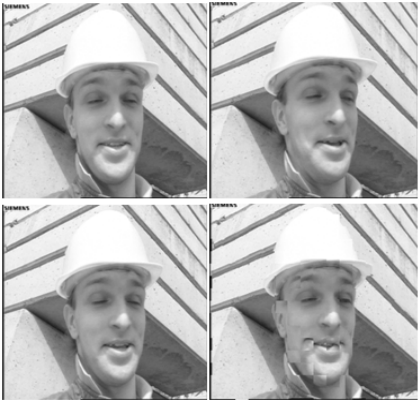

**Figure 2.** Frame $I_{n-1}$ (Top Left), Frame $I_{n+1}$ (Top Right), Original WZ Frame or Frame $I_n$ (Bottom Left) and Generated Side Information Frame using our method (Bottom Right).

## 5. Method

### 5.1. Our New Motion Estimation

We modified the conventional motion estimation method with the usage of SSIM (and also CW-SSIM and VIF) as the similarity criterion for block matching. SSIM works with the image structures and thus has better performance than MSE in block matching [28]. Simple full-search block based motion estimation is applied but each time the quality criterion for the computation of the similarity between reference block and target block is changed. Five metrics have been used to see the trade-offs using each of them. Search range is a block with side equal to three times bigger than the side of the original block.

As we know, motion estimation is an important part of video coding. Computational complexity of the coding strongly depends on the algorithm used for motion estimation. It seems that (and simulation results confirm this fact) applying SSIM instead of MSE increases the complexity but improves the performance. There is a trade-off in selecting the similarity criterion in block matching. As we can see in results section, VIF shows the best performance but has the most complexity. CW-SSIM is better than SSIM in performance and worse in computational cost. It is again worthwhile mentioning that the complexity is not a major issue in this setting. It is assumed that the decoder has enough resources including computational power and speed to perform the decoding procedure. The main job in the DVC setting is to keep the complexity of the encoder as low as possible while keeping a relatively acceptable video quality.

### 5.2. SI Generation at the Decoder

A frame interpolation algorithm is used at the decoder side to predict the side information for DVC. Consider three frames numbered $I_{n-1}, I_n, I_{n+1}$. It is desired to obtain an estimate of frame $I_n$ from the known information namely decoded version of intra frames $\hat{I}_{n-1}$ and $\hat{I}_{n+1}$.

First, we employ forward motion estimation between frame $\hat{I}_{n-1}$ and frame $\hat{I}_{n+1}$. The output of this step is motion vectors for each block. A simple idea to generate SI data is to halve the motion vector for each block of the image, and then move that block from frame $\hat{I}_{n-1}$ by this half motion vector. Another idea is to construct the frame $\hat{I}_n$ by subtracting the coordinate points of the corresponding block in the $\hat{I}_{n+1}$ frame by the half of the motion vector. Fig. 3 demonstrates these ideas. We used the combination of these two ideas i.e. frame $\hat{I}_n$ is built by averaging a block obtained by adding the half motion vector to the reference block of frame $\hat{I}_{n-1}$ and block obtained by subtracting the half motion vector form the reference block of frame $\hat{I}_{n+1}$, so this is bi-directional motion compensation. If we show it symbolically, we can write:

$$Block_n = \frac{1}{2}([Block_{n-1} + \frac{MV}{2}] + [Block_{n+1} - \frac{MV}{2}]) \quad (8)$$

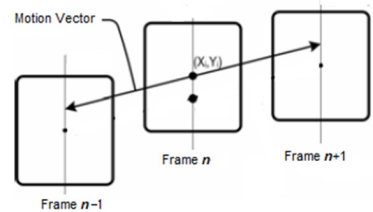

**Figure 3.** Demonstration of the interpolation process.

## 6. Simulations and Results

In this section we present the simulation results for the test sequence of "Foreman". The "Foreman" sequence consisted of 300 frames of size $144 \times 176$. The channel between the encoder and decoder was assumed to be error-free. Each sequence was first decomposed to two groups of frames. The odd frames were intra coded using a conventional H264 encoder and the even frames were Turbo encoded and their parity bits sent to the decoder. The base Turbo encoder consisted of two convolutional encoders with the constituent encoders having the generating polynomials as $G_1(D) = 1 + D^2 + D^3$ and $G_2(D) = 1 + D + D^2 + D^3$ and each having a feedback polynomial of $F(D) = 1 + D^2 + D^3$. The interleaver size was set to 1024 and the maximum iteration number for the APP decoders was set to 15.

First of all, to show the superiority of the motion estimation and side information generation process as proposed in this paper to the other currently available DVC structures, a method of comparison has been introduced in terms of bits needed for compression of the WZ frame. The comparison was performed in the pixel domain for simplicity. The frames were decomposed to their 8 bit planes each. It was assumed that no distortion other than the original quantization to 128 levels by the camera was allowed, meaning that practically Wyner-Ziv coding was reduced to Slepian-Wolf coding and no secondary quantization was allowed. The odd frames were thus encoded at rate $H(F_o)$ and the even frames were encoded at rate $H(F_e | F_o)$ and $F_e, F_o$ being even and odd frames bit planes respectively. As $H(F_o)$ will be the same for all tested methods, we only compared $H(F_e | F_o)$, meaning that the conditional rate of WZ frames - given that the base frames are entropy coded and are available without any further distortion at the decoder- was set as a comparison criterion. The results are given for five different methods. Each method has its own distortion measure utilized for motion compensation. Quality metrics that we used are: SAD, MSE, SSIM, CW-SSIM and VIF. Block size for motion estimation is N=16 (16×16 macro blocks). Table 1 shows the performance of the side information generation subsection with different distortion measures. The values of Tables 1 are driven from averaging over 20 frames from the video sequence. Each of these 20 averaged values corresponds to the similarity between the real frame and the frame obtained by the compression technique (with its own quality measure). This similarity is stated in various representations such as MSE, SSIM, VIF, bitplane error and $H(F_e | F_o)$. Average bitplane error is the Hamming distance between the real frame and reconstructed frame, when each of them is considered as a sequence of bits (pixels are 8-bit binary strings and the whole sequence can be built by putting these strings in serial order). We put this parameter (errors of bit planes) to emphasize this point that usage of the HVS based metrics reduces the error in the most significant bit planes and less error in left-sided bit-planes results in better performance in lossy and lossless coding.

It is noteworthy that the main parameter of comparison that shows the superiority of one method to another is $H(F_e | F_o)$. This parameter shows the conditional rate at which the WZ frames can be encoded. It is certain that in a practical setting using the Turbo codes, this rate might not be achievable exactly and it is the average rate of a limited number of frames from limited number of sequences but the value of the proposed method in this paper will be more appreciated with a closer look at Table 1. It can be seen in Table 1 that an obvious drop in the value of conditional rate happens for the SSIM criterion. CW-SSIM sees also a drop better than SSIM and VIF meets a little bigger drop than both. As Table 1 shows we can see that VIF shows the best performance in compression quality but has the biggest computational complexity. CW-SSIM shows better performance than SSIM but it is more complex than SSIM. Regarding the computational complexity and the coding performance obtained from simulation we can conclude that SSIM can be a knee point. Compression quality is good enough while complexity is not of much concern. But VIF shows the best performance and is valuable when complexity of the decoder is not important at all and some decoding delay can also be tolerated.

MSE has a little improvement over SAD but the improvement is around 0.01 bit which is not of practical importance because it might easily change for another set of data sequences.

It was also seen in the simulations that usage of very large block sizes (e.g. N=32) does not help much and is practically useless. It is noteworthy that using larger block sizes for SI generation process is different from that of block size used in DCT transform if the code is applied in the transform domain. The DCT size can be chosen $8 \times 8$ or $4 \times 4$ as in the usual sense and motion estimation in the SI generation process can be done using different block sizes as used in this paper.

After showing the theoretical (meaning the usage of $H(F_e | F_o)$ as a measure) superiority of the proposed method in terms of conditional rate and perceptual quality, the practical performance improvement by using this new method also needs to be shown. To compare the various methods, we used SSIM as a quality metric for comparison. The Rate-Quality performance is presented in Figure. 4. Also to better improve the Rate-Quality tradeoff, we used a HVS based DCT quantization table generation method in our design which we borrowed from [29]. Also, only the WZ bits have been counted for more clarity and to comply with the current standards, we used PSNR as a final comparison criterion. These figures confirm the usage of perceptual quality metrics in distributed video coding for side information generation. A gap of almost 1dB improvement is visible in Figure. 4 which can be obtained by using HVS based measures instead of MSE and SAD.

**Table 1.** Comparison of Different SI Generation Methods

| Quality Metric | Parameter | | | | | | |
|---|---|---|---|---|---|---|---|
| | MSE | SSIM | VIF | Average bitplane error | Four most significant bitplane errors | Simulation Time (Sec.) | $H(F_e | F_o)$ |
| SAD | 6.34 | 0.42 | 0.06 | 0.385 | [0.192, 0.247, 0.341, 0.417] | 45 | 0.926 |
| MSE | 6.28 | 0.43 | 0.08 | 0.380 | [0.179, 0.237, 0.338, 0.409] | 46 | 0.919 |
| SSIM | 25.89 | 0.79 | 0.45 | 0.285 | [0.043, 0.083, 0.207, 0.257] | 330 | 0.765 |
| CW-SSIM | 15.17 | 0.84 | 0.49 | 0.276 | [0.032, 0.070, 0.156, 0.234] | 1526 | 0.731 |
| VIF | 11.64 | 0.88 | 0.55 | 0.261 | [0.026, 0.061, 0.125, 0.215] | 1834 | 0.706 |

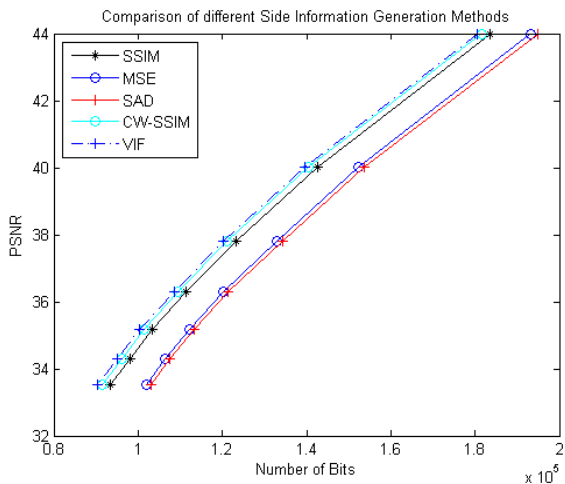

**Figure 4.** Quality vs. Number of Bits for the "Foreman" sequence with PSNR as the quality criterion.

## 7. Conclusions

In this paper we proposed a new side information generation technique for distributed video coding based on HVS based quality metrics. The new method was compared to other standard distortion measures such as MSE and SAD. It was shown via simulations that the proposed method is superior to current standard methods in the side information generation process and gains up to 1dB can be obtained. The new method has a higher complexity compared to MSE and SAD methods. This complexity can almost be ignored for DVC setting because the decoder is assumed to have abundant resources. For the cases where decoder has limited computational power, a trade-off between this gain and the complexity should be taken into account.